\documentstyle[prd,aps,epsf]{revtex}

\newcommand\be{\begin{equation}}
\newcommand\ee{\end{equation}}
\newcommand\bea{\begin{eqnarray}}
\newcommand\eea{\end{eqnarray}}
\newcommand\half{{\textstyle{1\over2}}}
\newcommand\quarter{{\textstyle{1\over4}}}
\newcommand{\smfrac}[2]{{\textstyle{{#1}\over{#2}}}}

\newcommand{\E}[1]{e^{\textstyle {#1}}}
\def\d{\partial}
\def\bmat{      \left |  \begin{array}{cc} }
\def\emat{ \end{array} \right |    }

\begin{document}
\draft
\title{Magnetic monopole loop for the Yang-Mills instanton}
\author{Richard C. Brower}
\address{Center for Theoretical Physics, MIT, Cambridge, MA 02139, USA}
\author{Kostas N. Orginos and Chung-I Tan}
\address{Department of Physics, Brown University, Providence, RI 02912}
\date{\today}
\maketitle
\begin{abstract}
 We investigate 't Hooft-Mandelstam  monopoles in QCD 
in the presence of a single classical instanton configuration.  The
solution to the Maximal Abelian projection is found to be a circular
monopole trajectory with radius $R$ centered on the instanton. At zero
loop radius, there is a marginally stable (or flat) direction for loop
formation to $O(R^4 log R )$.  We argue that loops will form, in the
semi-classical limit, due to small perturbations such as the dipole
interaction between instanton anti-instanton pairs.  As the instanton
gas becomes a liquid, the percolation of the monopole loops may
therefore provide a semi-classical precursor to the confinement
mechanism.
\end{abstract}
\pacs{MIT-CTP-2570, BROWN-HET-1041}

\section{Introduction}

Both instantons and magnetic monopoles are thought to play an
important role in the Euclidean description of the Yang-Mills
vacuum. In QCD instantons provide the solution to the $U(1)$
problem~\cite{tHooft2} and a vacuum composed of a liquid of
instantons~\cite{Shuryak} can explain chiral symmetry breaking and the
low mass spectrum, at least at a qualitative
level~\cite{Witten,Grandy}. Nonetheless, it is generally believed, as
conjectured by Mandelstam~\cite{Mand} and 't Hooft~\cite{tHooft1},
that {\bf magnetic monopoles, not instantons} are essential for
confinement in QCD and related gauge theories. For example, compact
QED with a lattice cutoff is known to be exactly dual to a Coulomb gas
of monopoles, which upon condensation causes confinement via a dual
Meisner effect~\cite{Peskin}. Similarly there is evidence for the role
of monopole condensation in the 3-d Yang-Mills-Higgs (or
Georgi-Glashow) model~\cite{Poly} and more recently in 4-d N = 2
supersymmetric Yang-Mills theory~\cite{Seiberg-Witten}.

Consequently, we are faced with a dilemma of two competing pictures of
the QCD vacuum (or more accurately the Euclidean equilibrium phase) as
either a coherent ensemble of instantons or a condensate of magnetic
monopoles.  In this paper, we show that these two pictures may in fact
be two descriptions for the same phase.  To establish a precise and
indisputable link between the monopole trajectories and instantons, we
define the monopole current in a field configuration via 't Hooft's
Maximal Abelian (MA) projection and demonstrate that in the background
of a single instanton the MA projection leads to a circular monopole
current loop of radius $R$ centered on an isolated instanton of
``width'' $\rho$. We also begin the study of monopole trajectories in
the presence of interacting instanton pairs.

Let us review briefly the attempts to identify monopole currents in
QCD. (The reader is referred to Polikarpov~\cite{PolikarpovRev} and
references therein for more details.) It is well known that the only
topological stable solutions to Euclidean Yang-Mills theory are the
multi-instanton solutions~\cite{BPST} with topological charge $Q$. On
the other hand, the 't Hooft-Polyakov monopole solution to 3-d
Yang-Mills-Higgs theory can also be viewed as a static solution to the
pure Yang-Mills Euclidean field equations in 4-d, where $A_0$ plays
exactly the same role as the adjoint Higgs field\footnote{
Essentially the same use is made of the adjoint field $A_0$ in the MA
projection that diagonalizes the Polyakov loop as discussed by
Suganuma et al.~\cite{Tanaka}, but we do not pursue this approach
further since it does not provide a Lorentz invariant definition of 
monopole variables.}
in the BPS limit. Of course, since the classical theory is now scale invariant
the monopole mass M is an arbitrary parameter.  Quantum effects will
be necessary to set the scale.  Subsequently Rossi \cite{Rossi} has
pointed out that this monopole configuration can be identified with
(i.e., has the same field configuration as) an infinite sequence of
instantons equally spaced by $M^{-1} 8 \pi^2/e^2$ on the time axis in
the limit $\rho \rightarrow \infty$.

Turning now to the MA projection, Chernodub and Gubarev~\cite{CG} have
recently shown that a straight line monopole trajectory through the
center of an instanton (or pair of instantons) satisfied the MA gauge
condition. However if one looks at the MA gauge fixing functional, $G
= \quarter \int d^4x A^+_\mu(x)A^{-}_\mu(x) $, $A^{\pm}_\mu\equiv
A^1_\mu\pm i A^2_\mu$, for the Chernodub-Gubarev solution, one notices
that it diverges at large distances. Since the proper definition of
the MA projection is not just a stationary point but a minimum of this
functional, this solution is still not a very satisfactory linkage
between an instanton and the monopole. The basic problem is that the
instanton is restricted to a space-time region whereas the static
monopole solution has an infinite trajectory.  In contrast we have
found that our monopole loop solution to the MA projection, which is
localized at the instanton, does give a finite value to $G$, which
drops gradually to an absolute minimum as the radius decreases: $G
\simeq 4 \pi^2\rho^2[ 1 + 4.6 (R/\rho)^4 log(\rho/R)]$. 
We then argue that this solution is easily stabilized by small
Gaussian perturbations or nearby interactions with anti-instantons. We
feel this is the first rather precise connection between the instanton
and monopole trajectories.

In addition to this continuum analysis, there has been some recent
numerical evidence in lattice simulations presented by Hart and Teper
\cite{Hart}, by Bornyakov and Schierholz \cite{Schierh}, and by
Markum et al. and Thurner et al. \cite{Markum} that shows that
monopole loops are correlated with multi-instantons configurations.
These numerical results also indicate that a correlation between 
monopole condensation and the formation of a dense liquid of 
instantons are a property of the quantum vacuum.  However our 
immediate goal is a more modest one: establishing a clear kinematical
connection between monopoles and instantons in the semi-classical limit. 
While this limited exercise is unable to show that the {\it dynamics} 
of the full quantum theory is in some sense approximated by a 
monopole-instanton correlated background and that confinement results 
from the percolation of these objects, we feel that clarifying the 
kinematics  is an important first step in bringing these two
pictures together.

To go beyond the analytical study, we have also undertaken a numerical
MA projection in the background of an isolated instanton by minimizing
the functional $G$ discretized on a 4-d hypercubic lattice. The only
solutions we have found were identical to our analytical monopole loop
solution albeit stabilized to a fixed radius (in lattice units) by
lattice artifacts.  We have also begun to study the monopole loops in
the background of an interacting instanton anti-instanton (I-A) pair
and found that the individual loops appear to be stabilized by the
interaction, if the I-A molecule is oriented so that the dipole interaction
is attractive. At a critical separation of the I-A pair, the
individual monopole loops fuse into a single loop. Details of the
multi-instanton study are left to a future publication~\cite{BOT}.
Nonetheless on the basis of our preliminary analysis, we conjecture
that the I-A interaction provides a semi-classical mechanism whereby
the monopoles are ``liberated'' from the individual instantons, and
which may therefore represent a precursor to condensation and
confinement.

The organization of the paper is as  follows. We begin by
noting that the Maximal Abelian projection, which is usually presented
as a gauge fixing procedure is equivalent to the introduction of an
``auxiliary Higgs'' field $\Phi =
\vec \phi \cdot \vec \tau$, which is determined by minimizing the
Higgs kinetic term,
\be
G = {\quarter} \int d^4x  [D_\mu(A) \vec\phi]^2 ,
\ee
(The Higgs field is related to the usual gauge rotation by $\Phi =
\Omega^\dagger \tau_3 \Omega$.) This approach has 
several advantages in terms of analysis and a clearer relation to
the standard discussion of monopole topology. (See Appendix A for details.)

Next in Section 2, we solve the stationarity equations for the Higgs
field (or gauge rotation in the $SU(2)/U(1)$ coset space), to find the
monopole loop solution for fixed radius $R$. Monopole loops in Abelian
gauge theories are reviewed to guide the construction.

In Section 3, we consider the full manifold of solutions and their
collective co-ordinates. We note that there is an interesting
correlation between the average over the orientations of all monopole
loops at fixed $R$ and the isospin orientation of the instanton given
by
\be
< N_{\mu\nu}> =\half[ (\delta_{\mu,3} \delta_{\nu,4} - \delta_{\mu,3}
\delta_{\nu,4}) - (\delta_{\mu,1} \delta_{\nu,2} - \delta_{\mu,2}
\delta_{\nu,1})]
\ee
where $N_{\mu\nu}$ is the skew symmetric unit tensor defining the plane of
the loop in 4-d.

In Section 4, we report on our numerical solutions leading to a preliminary
picture of the role of instanton anti-instanton (I-A) interactions
for loop stabilization at large separation and  for the I-A pair
loop fusion and percolation at small separations.

Finally in Section 5, we discuss our results and suggest future lines
of investigation. We comment on the naturalness of the
instanton-monopole connection. The loop ``VEV'' naturally breaks the
$SO(4)$ group down to the coset direct product $SU_L(2)/U_L(1) \times
SU_R(2)/U_R(1)$ in close analogy to isospin breaking in the 't
Hooft-Polyakov monopole construction.  Also we note that the Abelian
projection for the instanton in the singular gauge spreads the
singularity at the origin to the trajectory of the monopole loop,
directly relating the instanton's topological charge $Q$ to the
monopole charge $g$ by the identity $Q = e g/ 4 \pi$.  In summary we
conclude that monopole loops are peculiarly well matched to the
instanton, leading us to hope that there is a deeper connection in
confining gauge theories transcending our particular construction.

\section{ Monopole Loop Solution  in the Instanton Background}

The Maximal Abelian projection was introduced by 't
Hooft~\cite{tHooft1} in order to define magnetic monopole coordinates
by a partial gauge fixing procedure that leaves the maximal Abelian
subgroup free. In analogy with the Higgs system, 't Hooft suggested
introducing an adjoint field $X$, which by the gauge transformation,
$X \rightarrow X^\Omega = \Omega X
\Omega^{\dagger}$, is diagonalized to fix the gauge in the coset space
$SU(N)/U(1)^{N-1}$ up to $U(1)$ factors. Exceptional space-time
trajectories where $X$ has degenerate eigenvalues represent the
Abelian monopole configurations.  Various examples for X were
suggested such as $F_{12}$ or the (untraced) Polyakov loop, etc.

Subsequently a particular Lorentz covariant gauge has proven to best
correlate the monopoles condensate with confinement in lattice
simulations \cite{KSW}. This gauge is now referred to as { the}
Maximal Abelian (MA) gauge. For $SU(2)$, it is expressed as the
minimization of the functional,
\be
G[A_\mu] = \quarter \int  d^4x 
A^+_\mu(x)A^{-}_\mu(x)\; ,
\label{eq:global}
\ee
$A^{\pm}_\mu\equiv A^1_\mu\pm i A^2_\mu$. In differential form the MA
gauge condition becomes
\be
\Delta^\pm(x)\equiv (\d_\mu  \pm i e A^3_\mu ) A^\pm_\mu = 0 \; ,
\label{eq:local}
\ee
with the additional stipulation to avoid Gribov copies that one should
find the solutions corresponding to the global minimum of $G$. (We
will sometime refer to the stationarity condition~(\ref{eq:local}) by
itself as the ``differential'' MA projection.)

Before presenting the details of our monopole loop solution, it is
possible (at least with hindsight) to see why magnetic loops might
appear in an instanton background. Consider the field configuration
for a single instanton~\cite{Rajaraman,Belavin,Jakiw} in the singular
gauge,
\be
A_\mu \equiv \half \vec A_\mu\cdot\vec\tau=\smfrac{e}{2} \tau^a
\bar\eta^a_{\mu\nu} \d_\nu ln( 1 +
\rho^2/x^2), 
\label{eq:singular}
\ee
where the anti-self-dual 't Hooft symbol is $\bar\eta^a_{\mu\nu}\equiv
(i/2)Tr(\tau^a \tau_\nu \tau_\mu^\dagger)$ and $\tau_\mu = (i,\vec
\tau)$ with Latin indices restricted to
the vector components.  First one notices that this field
configuration already satisfies the differential MA (as well as the
Lorentz) gauge condition. Also it gives a finite value,
\be
G[A_\mu] = 2 e^2\int d^4x \frac{\rho^4}{x^2(x^2 + \rho^2)^2} = 4
\pi^2e^2
\rho^2 ,
\ee
to the MA gauge fixing functional --- the dominant contribution to $G$
coming from the core of the instanton, $0\leq x\leq \rho$. On the
other hand, the instanton in the non-singular gauge also satisfies the
differential Eq.~(\ref{eq:local}) for the MA projection, but now the
functional $G$ diverges logarithmically at large distances,
\be
G[A^{(ns)}_\mu] = 2e^2 \int d^4x \frac{\rho^2}{(x^2 + \rho^2)^2}
\rightarrow \infty.
\ee
Nonetheless, the non-singular gauge does reduce the contribution to
$G$ in the core of the instanton. Consequently, comparison between the
singular and non-singular gauges suggests that it {\em might} be
possible to find an intermediate solution which minimizes $G$ by a
gauge transformation such that the central region, inside some radius
($x \le R$), is converted to the non-singular gauge, while the large
distance behavior is unchanged from the singular gauge
configuration. In essence this will turn out to be the way we
construct the monopole loop solution.

In addition, by a general argument one can understand why this
construction might lead to monopole singularities.  No local gauge
transformation on the instanton can change the total topological charge.
However if we express the topological charge for the instanton in the
singular gauge via Gauss' law, $Q = \int d^4x \d_\mu K_\mu$,  in terms
of the flux for the gauge variant current, $$K_\mu = \frac{1}{8 \pi^2}
\epsilon_{\mu \nu \rho
\lambda} Tr[ A_\nu (\d_\rho A_\lambda -  
\smfrac{2i}{3} A_\rho A_\lambda)] \;, $$
one knows that the entire contribution comes from the singular point
at $x = 0$. Any gauge rotation that removes this singularity but does
not change the gauge at infinity must replace it with another
singularity at some finite distance.  As we will explain in more
detail in the conclusion, our monopole loop current is the source of
this singularity and its magnetic charge can therefore be directly
related to the topological charge of the instanton.

\subsection{General Equations}

In searching for solutions to the MA projection, we have found it more
convenient to use a gauge covariant formulation in terms of an
auxiliary Higgs-like field, $\Phi(x) \equiv
\Omega^\dagger \tau_3 \Omega$, instead of working directly with the  gauge
transformation $\Omega(x)$ itself (see Appendix for details).  In terms of the
isovector field
$\vec
\phi$,
$\Phi\equiv
\vec\phi\cdot\vec\tau$, the functional $G$ becomes
\be
G = \half \int d^4x\{ \half [D_\mu(A) \vec\phi]^2 + V(\vec\phi^2)\},
\label{eq:covglobal}
\ee
where $D_\mu\phi^\alpha\equiv
\d_\mu\phi^\alpha+e\epsilon^{\alpha\beta\gamma}A_\mu^\beta\phi^\gamma$
and the potential is $V(\vec\phi^2)=\sigma (\vec\phi^2-1)$ with
a Lagrange multiplier $\sigma$.  (One can see by inspection that
Eq.~(\ref{eq:covglobal}) is identical to Eq.~(\ref{eq:global}), after
the substitution $\Phi \equiv \Omega^\dagger \tau_3 \Omega$.)  It
is also natural to generalize the MA projection with a quartic Higgs
potential,
\be
V(\vec\phi^2)= \quarter \lambda (\vec \phi^2 - v^2)^2, 
\label{Genglobal}
\ee
with $\Phi(x)$ now given by $\vec \phi(x)\cdot
\vec\tau/|\vec\phi(x)|$.  We will refer to this more general
from as the ``Higgs MA projection''. The conventional MA projection is
recovered in the limit $\lambda \rightarrow \infty$ at fixed VEV $v$
in which the Higgs mass scale $m_H \rightarrow \infty$ as well.  Not
only does the covariant formulation have numerous technical
advantages, it shows that the MA projection need not be seen as a
gauge fixing prescription but instead as a gauge covariant method of
identifying the appropriate magnetic variables.

The general  problem is to construct solutions to the differential
form of the MA projection,
\be
D_\mu(A)^2\vec \phi - \vec \nabla_{\phi} V(\phi)  = 0 \; ,
\ee
in the background of a single instanton. For the standard
MA projection, this yields a linear PDE for
$\vec\phi(x)$,
\be
\d_{\mu}^2\vec\phi+2\vec A_{\mu}\times \d_{\mu}\vec\phi+\vec A_{\mu}\times 
(\vec A_{\mu}\times \vec \phi)-\sigma \vec\phi=0,
\label{eq:MAgauge}
\ee
subject to the constraint $\vec\phi^2=1$. For an instanton in
the singular gauge (\ref{eq:singular}), the potential also can be
written as $ A_\mu =-i x^2f(x)g^\dagger(x)\d_\mu g(x) $, where
$g^\dagger (x)=-i x_\mu\tau_\mu/|x|$, $\tau_\mu=(\vec \tau, i)$,
\be
 f(x) =  \frac{1}{x^2}\frac{\rho^2}{x^2+\rho^2},
\ee
and 
we have also set the gauge coupling $e=1$. 

Solving this equation is very similar to the problem, first considered
by 't Hooft~\cite{tHooft2}, for computing the Gaussian fluctuations of
gauge, fermionic and Higgs fields in the background of an instanton.
In studying self-dual solutions and their supersymmetric extensions,
it is useful to introduce tensors that reflect the chiral $SU_L(2)
\times SU_R(2)$ decomposition of the $O(4)$ Euclidean Lorentz group,
which suggests the conformal coordinates, ${\tt u} = x + i y = u \E{i
\varphi}$ and ${\tt v} = z + i t = v \E{i \psi}$, that enter into the
bispinor,
\be
x_{\alpha \dot \beta}
\equiv x_\mu (\tau_\mu)_{\alpha \dot \beta} = 
\pmatrix{
 z + i t  & x - i y \cr x + i y & -z + i t  \cr}=\pmatrix{
{\tt v}  &{\tt u}^{*} \cr {\tt u} & -{\tt v}^{*}  \cr
}
\label{eq:Bipolar}
\ee
for the singular gauge transformation, $g^{\dagger}(x)_{\alpha \dot
\beta} = -ix_{\alpha \dot \beta}/|x|$.  This notation, as we will see
in Sec. 3, is also convenient for exhibiting the Lorentz
symmetries of the monopole solution.

It follows from the property of $\bar \eta^\alpha_{\mu\nu}$ that
isospin components of $\vec A_\mu$ are orthogonal, i.e., $A_\mu^\alpha
A_\mu^\beta\propto \delta_{\alpha\beta}$. As a consequence, the last
two terms in Eq.~(\ref{eq:MAgauge}) both point along $\vec \phi$ in
the isospin space.  Let us parameterize this ``radial direction" at
each $x$ by a unit-vector,
\be
\hat \phi(x) \equiv  (\sin\beta\cos\alpha, \sin\beta\sin\alpha, \cos\beta)
\label{eq:radialvector}
\ee
and  introduce two unit tangent-vectors orthogonal to $\hat \phi(x)$,
$\hat\alpha(x)\equiv (-\sin\alpha,
\cos
\alpha, 0)$ and 
$\hat\beta(x)$ $\equiv (\cos\beta\cos\alpha,\cos\beta\sin\alpha,
-\sin\beta)$. By projecting Eq.~(\ref
{eq:MAgauge}) onto the $\hat\alpha(x)$ and $\hat \beta(x)$ axes, the
two independent differential equations for the MA gauge are
\bea
\sin\beta \; \d^2_\mu \alpha +2\cos \beta \; (\d_\mu\alpha)(\d_\mu\beta) 
= -2\sin\beta ( \hat \phi \cdot \vec A_{\mu})\; ( \d_\mu\beta),
\label{eq:MAalpha}\\ 
\d^2_\mu \beta -\half \sin(2\beta) \; (\d_\mu\alpha)^2 
= 2\sin\beta (\hat \phi \cdot \vec A_{\mu}) \; (\d_\mu\alpha ),
\label{eq:MAbeta}
\eea
respectively.  (The more general equations for the Higgs MA projection
for arbitrary $\lambda$ are given in Eqs. (\ref{eq:MAGenerala} -
\ref{eq:MAGeneralc}).)  Our task is to find solutions to these
equations.

For the explicit monopole content of a solution, it is instructive to
return to the active transformation, Eq.~(\ref{eq:Decomp}), and to 
decompose the transformed field into two parts,
Eq.~(\ref{eq:DecompTwo}), $A^\Omega_\mu(x) = \bar A_\mu(x) \; +
\;M_\mu(x)
\;$ where 
$$
\bar A_\mu(x) \equiv \Omega(x) A_\mu(x)\Omega^\dagger(x) \;\;,\;\;
M_\mu(x) \equiv \frac{1}{ i e}  \Omega(x) \d_\mu \Omega^\dagger(x) \; .  
$$
The gauge rotation in terms of three
Euler's angles is 
\be
\Omega(x) = \E{ i \omega(x)} = \E{i\gamma \tau_3/2} \E{ i
\beta \tau_2/2} \E{ i \alpha \tau_3/2}.
\ee
One can readily verify that this is consistent with
Eq.~(\ref{eq:radialvector}) and, owing to the residual $U(1)$
invariance, $\vec\phi(x)$ is independent of the angle $\gamma$.
Focusing on the Abelian field $a_\mu(x)\equiv
Tr(\tau_3A_\mu^\Omega)=\bar A_\mu^3+M_{\mu}^{3}$, one observes
that the first term  $\bar A_\mu^3=\hat \phi\cdot \vec A_\mu $  provides the
source for our PDE's, Eqs.~(\ref{eq:MAalpha}) and (\ref{eq:MAbeta}).
The induced term,
\be
M_{\mu}^{3}(x) =-\frac{1}{e}[\cos\beta(x)\d_\mu \alpha(x)+\d_\mu\gamma(x)],
\label{eq:MthreeGeneral} 
\ee
will give rise to topological current for the monopoles,
Eq.~(\ref{eq:MAcurrent}). In terms of the Euler's angles, the magnetic
current is explicitly given as,
\be
k_\mu(x) = \frac{1}{2\pi e }\epsilon_{\mu\nu \rho \sigma}  \d_\nu  
[\d_\sigma\alpha(x)\d_\rho\cos\beta(x)].
\label{eq:MACurrentAB}
\ee
Note that $M^3_\mu$ is not specified uniquely due to the residual $U(1)$
gauge symmetry. Generally, we will choose $\gamma(x)=-\alpha(x)$ so that
the Dirac sheet associated with the monopole loop solutions can be
oriented conveniently and so that
\be
\Omega(x) = \cos(\beta/2) +  i  \sin(\beta/2) [ (\cos\alpha)\tau_1 - 
(\sin\alpha )\tau_2]
\ee
obeys the boundary condition $\Omega\rightarrow 1$ as $\beta\rightarrow 0$.

\subsection{Monopole Loop in $U(1)$ Gauge Theory}

To construct monopole loop solutions, we have found it helpful to work
``backward'' from specific examples of $U(1)$ monopole currents. First,
recall Dirac's construction for a {\it static} monopole of magnetic
charge $g$ at the origin $x = y = z = 0$, with magnetic current,
$k_\mu(x)=g\delta^{(3)}(\vec x)\delta_{\mu 4}$. The field is given by
\be
a_{\mu}=\frac{g}{4\pi}[1- \cos \vartheta]\d_\mu \varphi,
\label{eq:Uonepole}
\ee
in spherical coordinates, $0\leq \vartheta\equiv
\tan^{-1}(\sqrt{x^2 + y^2}/z)\leq \pi$ and $\varphi = \tan^{-1}(y/x)$.
The singular behavior due to $\d_{\mu}\varphi=\hat \varphi /\sqrt{x^2
+ y^2}$ as $\sqrt{x^2 + y^2}\rightarrow 0$ corresponds to having
placed a Dirac string along the negative $z$-axis, which in 4-d leads
to a Dirac sheet in the left-half of the 3-4 plane.  The Dirac sheet
can be moved by a gauge transformation.

This construction is an example of a general formalism for writing
down a solution with a Dirac sheet attached to a closed Euclidean
monopole current loop~\cite{Dirac-Sheet}. Consider a unit monopole
trajectory $x_{\mu}=y_\mu(\tau)$ and its associated magnetic current
$k_\mu(x)=\int d\tau \d_\tau y_\mu (\tau) \delta^4(x-y)$. We can
attach to it a Dirac sheet described by $y_\mu(\tau,\sigma)$, where
$x_{\mu}=y_\mu(\tau,0)$.  The associated vector potential for the
monopole loop is
\be
a_\mu(x)=-\frac {1}{4\pi} \int d^4x' \frac{\d_\nu \tilde
G_{\nu\mu}(x')}{|x-x'|^2},
\label{eq:Diracloop}
\ee
where $\tilde
G_{\mu\nu}=\half\epsilon_{\mu\nu\rho\sigma}G_{\rho\sigma}$ and
$G_{\mu\nu}(x)=\int d\tau d\sigma
\{\d_\tau y_\mu\d_\sigma y_\nu-\d_\tau y_\nu\d_\sigma
y_\mu\}\delta^4(x-y(\tau,\sigma))$. 

\begin{figure}
$$
\epsfxsize=8.6cm
\epsfbox{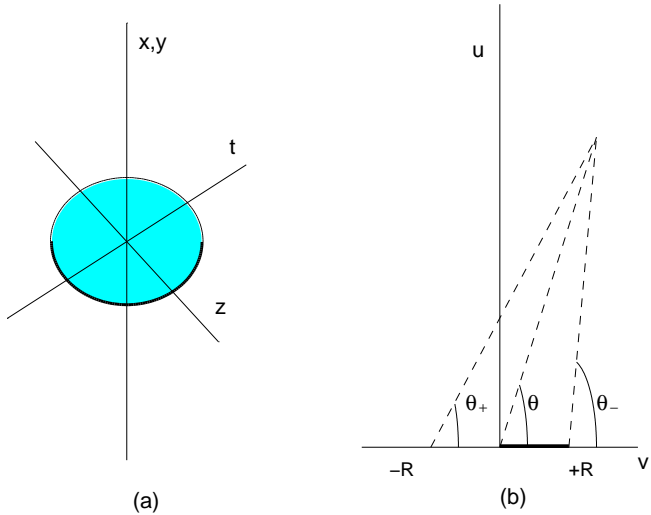}
$$
\vspace{0.1cm }
\caption{(a) The Dirac sheet for a monopole loop in QED. (b) $\theta_{\pm}$
variables for Dirac loops solution.
\label{fig:qedloop} }
\end{figure}

We now apply Eq. (\ref{eq:Diracloop}) to a monopole of charge $g$
moving in a closed loop of radius $R$ in the 3-4 plane centered at
the origin. The Dirac sheet (or solenoid world sheet) is chosen to run
across the loop (see Fig~\ref{fig:qedloop}), parameterized by
$y_1=y_2=0$, $y_3=R(1-\sigma)\cos \pi \tau$, $y_4=R(1-\sigma)\sin \pi
\tau$, where $-1\leq \tau<1$,  and $0\leq \sigma<1$.
It is convenient to  use the conformal  coordinates introduced earlier,
${\tt u}=x + i y = u \E{i \varphi}$ and ${\tt v}=z + i t = v \E{i \psi}$.
The magnetic current,
\be
k_\mu(x)=\frac {g}{ 2\pi}
\delta(x)\delta(y)\delta(v-R)
\hat\psi,
\ee
results in the  potential
\be
a_{\mu}=  \frac{g}{4\pi}[1-\cos(\theta_++\theta_-)]\partial_{\mu}\varphi,
\label{eq:Uoneloop}
\ee
where $\theta_{\pm}\equiv \tan^{-1}[u/(v\pm R)]$.  (See
 Fig~\ref{fig:qedloop} for a geometric interpretation for
 $\theta_{\pm}$.)

As for the case of a single monopole, $\partial_{\mu}\varphi=\hat\varphi/u$
is singular as one approaches the 3-4 plane, $u\rightarrow 0$. For
$v>R$, $\theta_{\pm}\rightarrow 0$, the pre-factor vanishes, and this
singularity is absent. However, for $0<v<R$, $\theta_+\rightarrow 0$,
$\theta_-\rightarrow \pi$, the pre-factor approaches 2 and a Dirac
sheet is present. We are now ready to generalize this
construction to the Maximal Abelian projection in the background of 
an instanton.

\subsection {Monopole Loop in an Instanton Background}

Returning to the differential MA conditions,
Eq.~(\ref{eq:MAalpha}) and Eq.~(\ref{eq:MAbeta}), an obvious solution
corresponds to the gauge rotation from the singular to the
non-singular gauge, $\Omega =g^{\dagger}= (x_4 + i \vec x \cdot \vec
\tau)/|x|$. In the Euler parameterization, this corresponds to $\alpha
= \varphi - \psi$, $\gamma=\pi-(\varphi+\psi)$, and $\beta = 2
\theta \equiv 2\tan^{-1}(u/v)$. However, on surfaces where $\d_\mu \alpha$ is
singular ($u = 0$ or $v=0$), $\cos\beta=0$ so that no magnetic
monopole is formed.\footnote{ The solution~\cite{CG} given by Chernodub and
Gubarev corresponds to $\beta =
\vartheta$ and $\alpha = \varphi$, where $\vartheta$ 
and $\varphi$ are the polar and azimuthal angles for the spatial three
vector $\vec x$. This is the standard static ``hedgehog''
configuration for $\phi^\alpha = x^\alpha/|\vec x|$, which gives rise
to the 't Hooft-Polyakov monopole with charge $4 \pi/e$ at $\vec x =
0$. This solution leads to a divergence value for the gauge fixing
functional $G$.  }

With the gauge choice $\gamma=-\alpha$, the induced potential,
$M^3_\mu$, becomes
\be
M_{\mu}^{3}(x) =\frac{1}{e}[1-\cos\beta(x)]\d_\mu \alpha(x).
\label{eq:Mthree} 
\ee
Note the similarity between this expression and the monopole solutions
for  QED, Eq.~(\ref{eq:Uoneloop}). The Dirac sheet is formed on the
surface where $\partial_{\mu}\alpha$ becomes singular and
$\beta(x)=\pm
\pi$. The magnetic current is  on the boundary of the Dirac sheet
where $\beta$ has a discontinuity. 

Consider the ansatz where $\alpha$ and $\beta$ are only functions of
$(\varphi, \psi)$ and $(u,v)$ respectively, i.e.,
$\alpha(\varphi,\psi)$ and $\beta(u,v)$. It can be shown that the
first of the two MA conditions, Eq.~(\ref{eq:MAalpha}), is solved
under this ansatz by
\be
\alpha=\varphi-\psi.
\label{eq:ansatz}
\ee
It follows from the QED example that a Dirac sheet can be present
either in the 1-2 plane, ($v\rightarrow 0$), or the 3-4 plane,
($u\rightarrow 0$), or both. It is sufficient for us to seek solutions
where the monopole loops are oriented in the 3-4 plane.  Other
orientations, including that oriented in the 1-2 plane, can be
obtained by performing appropriate $O(4)$ rotations, as we demonstrate
in Sec. 3.

Instead of $u$ and $v$, we convert to $x$ and $\theta$,
where $x^2=u^2+v^2$ and $\theta=\tan^{-1}(u/v)$, $0\leq
\theta\leq
\pi/2$. (See Section 4 for further discussion of the
ansatz in the u-v coordinate system.) Now Eq.~(\ref{eq:MAbeta}) becomes
\be
\frac{1}{x^{3}}\d_x (x^3\d_x\beta)
+\frac{1}{x^2\sin2\theta}\d_\theta(\sin2\theta\d_\theta\beta)-\frac{2\sin
2\beta}{x^2\sin^2(2\theta)}+4f(x)\sin^2\beta(\cot\beta-\cot 2\theta)=0.
\label{eq:B}
\ee
For $G$ to remain finite, one must impose the boundary condition
$\sin\beta=0$ at $\theta=0$ and $\theta=\pi/2$. There remains the
freedom for $\beta$ to take on values which are integral multiples of
$\pi$.

Recall that we are seeking solutions with $A^\Omega_\mu(x)$
approaching the behavior of the non-singular gauge at the origin, and
approaching the behavior of the singular gauge at infinity.  We
therefore expect a solution with $\beta
\simeq 2 \theta\> (mod\> \pi)$ for x small and with $\beta \rightarrow 0\>
(mod\> \pi)$ at infinity. A monopole loop lying in the 3-4 plane corresponds to
a solution where $\beta$ has a discontinuity on the positive
$v$-axis. To be precise, we seek solutions where $(i)$
$\beta(x,\pi/2)=0$, $(ii)$ $\beta(x,0)=-\pi $ for $0<x<R$, and
$\beta(x,0)=0 $ for $R<x<\infty$, and $(iii)$ $\beta(x,\theta)
\rightarrow 0$ for $x\rightarrow \infty$. 

\begin{figure}
$$
\epsfxsize=8.6cm
\epsfbox{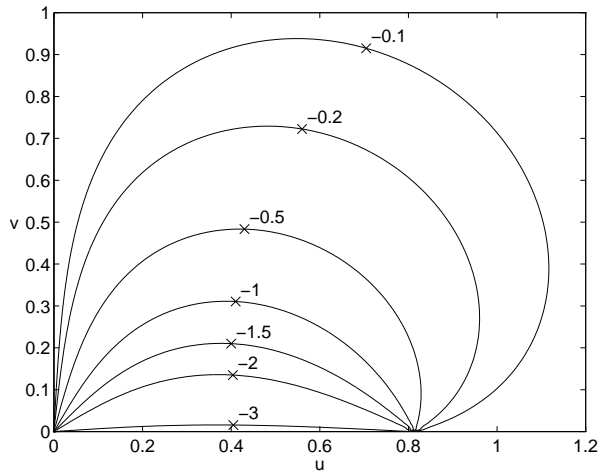} 
$$
\vspace{.1cm }
\caption{Solution to $\beta(u,v)$ for a monopole
loop with $R/\rho = 0.81$ and $\rho = 1$.
\label{fig:betasol} }
\end{figure}

An analytic solution for $\beta(x,\theta)$ in the limit of small
monopole size, $R/\rho \rightarrow 0$ can be found; this will be
discussed in Section 3. For general $R/\rho$, the solution to
Eq.~(\ref{eq:B}) can be obtained numerically.  Using a simple
relaxation method for $G$ (described in Section 4), very accurate
solution for $\beta$ can be found, as illustrated in
Fig~\ref{fig:betasol}. Independent of the detailed form of the
solution, with $\alpha=\varphi-\psi$, the discontinuity in
$\beta(x,\theta=0)$ at $x=R$ leads to a magnetic current,
\be
k_\mu(x)=\frac{2}{e} \; \delta(x_1)\delta(x_2){\delta(v-R)}
\hat\psi,
\ee
where we have re-introduced the charge factor. Therefore our solution
corresponds to a loop of monopole with a magnetic charge $g=4\pi/ e$ in the
instanton background.

In fact we have explored the more general case of a Higgs potential
$V(\phi) = \lambda (\phi^2 - 1)^2$, where $\phi$ is the magnitude of
the Higgs field $\vec\phi$.  The change in the value of $G$ is almost
independent of the parameter $\lambda$. It is now clear that our
general monopole loop solutions yield a family of solutions which all
satisfy the local MA condition and which interpolate between the
singular gauge and the non-singular gauge for an instanton. For $0\leq
x << R $, $\Omega(x)\sim g^{\dagger}(x)$ up to a $U(1)$ rotation; the
effect of the gauge transformation is to remove the singularity at
$x=0$. Since $\beta\rightarrow 0$ for $x\rightarrow \infty$, it
follows that, with $\gamma=-\alpha$, $\Omega\rightarrow I$ rapidly and
the large-$x$ behavior of the gauge fields is unchanged.  The
transition between ``small'' $x$ and ``large'' $x$ is marked by the
presence of a monopole loop of radius $R$ in the 3-4 plane.  As $R$
goes to $\infty$, the monopole loop is pushed to infinity, leaving
behind a gauge field configuration which is the instanton in the
non-singular gauge.

There is in fact a very large manifold of solutions. In addition to
the loops described above, one can clearly impose the boundary
conditions for $\beta$ to jump by $\pm \pi$ on either the $\theta = 0$
axis or the $\theta = \pi/2$ axis or both, at arbitrary radii
$R_i$. So long as the final value of $\beta$ as $x^2 \rightarrow
\infty$ is set to be the same multiple of $\pi$ on both boundaries, no
"topological charge has been pushed to the spatial infinity and the
functional $G$ will still be finite. This will give a series of
concentric loops with increasing radii and magnetic charge either $g$
or $-g$. In addition as describe in Sec 4, the application of Lorentz
invariance to these solutions will re-orient these loops in 4-d space.

\section{Collective Coordinates and Stability of Monopole Loop}

In this section, we first address the question concerning the
degeneracy of the monopole loop solutions constrained to a fixed
radius $R$ in the instanton background.  This analysis closely
parallels the discussion of the collective coordinates for the
instanton itself. The Yang-Mills action is invariant under the
15-parameter conformal and the 3-parameter global $SU(2)$
groups. However, an instanton solution breaks 8 of these symmetries
leading to 8 collective coordinates, roughly identified as 4 for its
location, 1 for its size, and 3 for its isospin orientation, (see
below for a more precise definition.) A monopole solution in each
instanton background with these 8 coordinates held fixed further
breaks some of the remaining 10 symmetries, leading to additional
collective coordinates for the orientation of the loop.

Next we will address the dependence of the loop solution on the loop
radius $R$. Each solution is a stationary point of the gauge fixing
functional under the constraint of fixed radius.  To restrict further
the MA projection, one may also impose the condition that $G$ is a
global minimum.  We therefore need to study $G(R)$, $0\leq R <\infty$,
and determine if there is a true minimum for a finite non-vanishing
value of loop radius $R$.  We shall show that, for a large instanton
with size $\rho$, (or equivalently small monopole loops with radii
$R<< \rho$), there is a very weak dependence of the gauge fixing
functional $G$ in $R$, so the scale of the loops is ``nearly'' a
collective coordinate. This near zero mode leads to ``marginal
instability'' for the formation of small monopole loops.

\subsection{$O(4)$ Invariance and Orientation of the Monopole loop}

The fact that we constructed the monopole loop solution lying in the
3-4 plane is purely for mathematical convenience; a larger family of
loop solutions can be obtained by applying $O(4)$ Lorentz
transformations.  Since $SO(4)=SU_L(2)\times SU_R(2)$, it is possible
to define two sets of mutually commuting angular momentum generators
by $L_L^a=-(i/2)\bar\eta^a_{\mu\nu}x^{\mu}\d_\nu$ and $
L_R^a=-(i/2)\eta^a_{\mu\nu}x^{\mu}\d_\nu$. (In terms of the
conventional ``rotation'' and ``boost'' operators, $\vec J$ and $\vec
K$, one has $ \vec L_L=[\vec J- \vec K]/2$ and $ \vec L_R=[ \vec J+
\vec K]/2$.) Using the spinor basis, it is trivial to show that an instanton
configuration is invariant under an arbitrary rotation in $SU_R(2)$,
generated by $\vec L_R$.  An instanton is also invariant under an
$SU_L(2)$ rotation, (generated by $ \vec L_L$), provided that a
corresponding isospin rotation is performed simultaneously.

If we consider the kinematics of the loop in the 3-4 plane, ignoring
the instanton background for now, clearly rotations in the 1-2 and 3-4
plane leave it invariant.  In fact it can be easily shown that these
are the only invariances and that the remaining four-parameter coset
space $SU_L(2)/U_L(1) \times SU_R(2)/U_R(1)$ rotates the loop to an
arbitrary plane. Thus the loop breaks both left and right chiral
$SU(2)$ factors in the $O(4)$ group, analogous to the way the 't
Hooft-Polyakov monopole breaks the $SU(2)$ isospin group. Since a MA
gauge also fixes a direction in the isospin space by identifying
$\tau_3$, the only remaining symmetry transformations belong to
$SU_R(2)$. Thus given a monopole loop solution, all other inequivalent
solutions can be obtained by performing $O(4)$ rotations belonging to
the quotient space $SU_R(2)/U_R(1)$.

We now provide a few details on these transformations.  Any rotation
$U\epsilon SU_R(2)$, expressed in the conformal coordinates of
Eq.(\ref{eq:Bipolar}), acts on $x_{\alpha \dot \beta}$ to give
$x'_{\alpha \dot
\beta} = x_{\alpha \dot \beta'} U^{\dagger}_{\dot \beta'\dot
\beta}$. Let us see what happens to
a circular loop of radius $R$ centered in the 3-4 plane: ${\tt
u}_0(\sigma)=x_0(\sigma)+iy_0(\sigma)=0$ and ${\tt
v}_0(\sigma)=z_0(\sigma)+it_0(\sigma)=R e^{ i\sigma/2}$, where $0\leq
\sigma\le 4\pi$. First consider a rotation, $R_3(\lambda)=e^{i\lambda
L_R^3}=e^{i\lambda\tau_3/2}$, in the $U(1)$ subgroup of $SU(2)_R$. It
simultaneously rotates the 1-2 and 3-4 planes by the same angle
$\lambda/2$, {\it i.e.}, ${\tt u}_0\rightarrow {\tt u}(\sigma)={\tt
u}_0(\sigma)e^{-i\lambda/2}$ and ${\tt v}_0 \rightarrow {\tt
v}(\sigma) = {\tt v}_0(\sigma) e^{-i\lambda/2}$. This clearly leaves a
monopole loop lying in the 3-4 plane invariant.
Next consider rotations, $R_2(\lambda) = e^{i\lambda L_R^2} =
e^{i\lambda\tau_2/2}$. Again using conformal coordinates, one finds
that
\bea
{\tt u}(\sigma) &=& \cos (\lambda/2){\tt u}_0(\sigma) -
\sin(\lambda/2) {\tt v}_0^{*}(\sigma) =-\sin (\lambda/2)R e^{-i
\sigma/2} \nonumber \\
{\tt v}(\sigma) &=& \cos (\lambda/2) {\tt
v}_0(\sigma)+\sin(\lambda/2){\tt u}_0^{*}(\sigma) =\cos
(\lambda/2)R e^{i \sigma/2}.\nonumber
\eea
The resulting loop has a circular projections onto the 1-2 and 3-4
planes with radii $R\sin(\lambda/2)$ and $R\cos(\lambda/2)$
respectively. In particular, for $\lambda=\pi$, it rotates a loop in
the 3-4 plane to one lying in the 1-2 plane, as promised.
Finally, consider a general $SU_R(2)$ rotation. Since it can be
parameterized using Euler's angles as $U(\lambda_3,
\lambda_2,\lambda_3')=R_3(\lambda_3)R_2(\lambda_2)R_3(\lambda_3')$, one
finds that the resulting loop is given by
$$
{\tt u}(\sigma) = -
R\sin(\lambda_2/2)e^{-i(\sigma+\lambda_3-\lambda'_3)/2},
\>\>\>\> {\tt v}(\sigma)=R\cos (\lambda_2/2) 
e^{i(\sigma-\lambda_3-\lambda'_3)/2}.
$$
For $\lambda_3\not= 0$, although the projections onto the 1-2 and 3-4
planes remain circular, it is no longer circular for projections onto
other planes, e.g., the 1-4 plane. Clearly, all distinct loops can be
characterized by two independent angles, parameterized by $\lambda_3$
and $\lambda_2$.

Given a fixed isospin orientation for an instanton, the average over
the loop orientation tensor $N_{\mu\nu}$ can now be found. For
instance, for the standard isospin orientation given by
Eq.~(\ref{eq:singular}), the monopole loop solution lying in the 3-4
plane is characterized by an anti-symmetric tensor
$N_{\mu\nu}=(\delta_{\mu,3}
\delta_{\nu,4} -
\delta_{\mu,3} \delta_{\nu,4}) $. Averaging over $SU_R(2)$, one finds that 
\be
< N_{\mu\nu}> =\half[ (\delta_{\mu,3} \delta_{\nu,4} - \delta_{\mu,3}
\delta_{\nu,4}) - (\delta_{\mu,1} \delta_{\nu,2} - \delta_{\mu,2}
\delta_{\nu,1})].
\ee
In a random dilute instanton gas, this implies a correlation between
the monopole loops and the isospin orientation of the associated
instanton. As discussed in the Conclusion this correlation
may provide a signature of our mechanism for loop formation in
the QCD vacuum.

\subsection{ Limit of Small Monopole Loop}

Classically $SU(2)$ gauge theory in four-dimension has no dimensionful
parameter, so the instanton solution  breaks scale
invariance through the introduction of the width $\rho$. Consequently
our monopole solutions depends on the dimensionless ratio
$\rho/R$, and  we may consider the
solutions in the limit for small loops size, $\rho/R\rightarrow
\infty$. Let us return to Eq.~(\ref{eq:B}), 
where the width parameter $\rho$ enters through the function
$f(x)=\rho^2/[x^2(x^2+\rho^2)]$. After scaling both $x$ and $\rho$ by
$R$, one finds that the equation is greatly simplified in the limit
$\rho/R\rightarrow \infty$, $f(x)\rightarrow x^{-2}$.

In this limit, an exact monopole loop solution to Eq.~(\ref{eq:B}) is
\be 
\beta(x,\theta)=\beta_0(x,\theta)\equiv 2\theta-(\theta_++\theta_-)+\pi ,
\ee
with $\theta_{\pm}=\tan^{-1}[u/(v\pm R)]$, as defined for the earlier
QED example with a monopole loop in the 3-4 plane.  Note that
$\beta_0=0$ on the u-axis, and it has a jump by $\pi$ on the v-axis at
$v=R$, ($\beta_0=-\pi$ for $0<v<R$ and $\beta_0=0$ for $R<v<\infty$).
This solution is valid provided that $0<R<<\rho$. The Dirac sheet
lies in the 3-4 plane bounded by the circle of radius $R$ where the
monopole current resides.

In the limit of small monopole loop and for $0\leq x<< R$, the
function $\beta \simeq 2\theta -\pi$ and therefore $\beta$ is not
``small''. This reflects the fact that $\Omega\simeq g^{\dagger}(x)$
in the limit $x\rightarrow 0$, so it is singular at the origin. On the
other hand, outside of the monopole loop radius, $R<x<\infty$, $\beta$
scales with $R$. Therefore in this ``outer region", for $R\rightarrow
0$, $\beta$ admits an expansion in $(R^2/x^2)$ as
\be
\tan\beta\sim \tan\beta_0=\frac{-R^2\sin (2\theta )}{
x^2-R^2\cos (2\theta)}\sim -\sin (2\theta) (\frac {R^2}{x^2})+0(\frac
{R^4}{x^4}).
\label{eq:smallR}
\ee
Consequently for fixed $x$ and $x \not= 0$, the limit of small
monopole loop is characterized by ``small'' gauge
transformations. This allows us to examine the stability of monopole
solutions by a linear analysis about $\phi^a(x)\sim
\delta_{a,3}$.

\subsection{Marginal Stability of the Small Monopole}

Our monopole solutions are stationary values of $G$ constrained by the
boundary condition to a fixed radius $R$. We therefore proceed to study
the functional  $G(R)$ evaluated at the the loop solution in the range
$0\leq R <\infty$. Near $R\rightarrow \infty$, $G(R)$ is monotonically
increasing since $G$ is divergent for an instanton in the non-singular
gauge.  For $R$ small, a leading order monopole solution is known.
Initially we had hoped that $G(R)$ has a minimum for some fixed
$R>0$. However in spite of our best effort, variational calculations
have so far led to results where $G(R)$ is always monotonically
increasing.  This is also confirmed by very accurate numerical
integration of our 2-d PDE's as described in Sec. 4.  We are now
convinced that the strong MA projection defined by the global minimum
of the functional $G$ is simply the singular gauge itself, which can be
viewed as the limit of a monopole loop with radius shrinks to
zero. Using our small radius solutions for $\beta$, we are able to
analyze the small $R$ behavior for $G(R)$.

We now consider in detail the region near to $R=0$. Assuming that
$R=0$ is a minimum, one would expect that $G(R)\sim
\rho^2[\delta_0+\delta_1 (R/\rho)^2 +\cdots\cdots]$, with
$\delta_1>0$.  It came initially as a surprise that we found
$\delta_1=0$.  This suggests the possibility of a zero mode in the 
stability equation around the singular gauge.

Let us expand $G[\phi]$ to quadratic order in
the neighborhood of $\vec\phi_0=(0,0,1)$, parameterized by $\vec\phi\simeq
\vec\phi_0 + \vec\omega $, 
\be
G[\phi] \simeq G[\phi_0] + \quarter \int d^4x  \vec \omega\cdot {\cal M} \vec
\omega, 
\ee
where the constraint $|\vec \phi |^2=1$ is realized to this order by having
$\vec\omega\cdot\vec\phi_0=0$, i.e., $\vec\omega$ only has 2 transverse
components.  The stability of small oscillations is  studied by finding
eigenvalues of a hermitian operator,
\be 
{\cal M} \omega_i= \lambda_i \omega_i,
\label{eq:eigenvector}
\ee
where 
\be
{\cal M}=-{\partial_x}^2 - {3\over x}{\partial_x} - {4\over
x^2}{\vec L_L}^2 -8 f(x) T^3 L_L^3 \; ,
\ee
with $\vec L_L$ as generators of $SU_L(2)$ and $\vec T$ of
isospin rotations in the adjoint representation. 

\begin{figure}
$$
\epsfxsize=8.6cm
\epsfbox{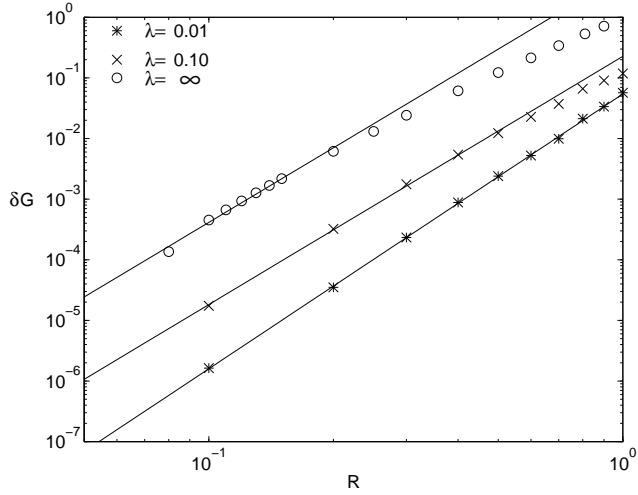} 
$$
\caption{ For $\lambda = 0.01,0.10,\infty$, the change in the Gauge 
fixing functional $G$ as a function of the monopole loop radius $R$.
\label{fig:R4beh} }
\end{figure}

Expanding $\vec \omega$ in terms of these normalized eigenvectors,
$\vec \omega =\sum_i\vec a_i\omega_i$, one has $G[\phi] \simeq
G[\phi_0] + \quarter \sum_i\lambda_i|\vec a_i|^2$. With $R=0$ as
a minimum for $G$, stability requires all eigenvalues are non-negative, i.e.,
$\lambda_i\geq 0$.  One would also expect that an infinitesimal loop
is ``nucleated'' along the direction $\vec \omega $ of the eigenvector
with the lowest eigenvalue, $\lambda_0$, and $\vec a_0=0(R)$. It
follows that $\delta_1 \propto \lambda_0$. The fact that our numerical
treatment indicates that $\delta_1\simeq 0$ as depicted in
Fig. \ref{fig:R4beh}, (described in Sec. 4) implies that $\lambda_0=0$,
i.e., to quadratic order there is a zero mode.  We now try to construct
this zero mode.

With $\vec\omega$ in the 1-2 plane, $T^3$ takes on eigenvalues
$t_3=\pm 1$.  Since ${\cal M}$ commutes with $\vec L_L^2$, $L_L^3$,
and $T^3$, there are a family of eigenvalues $\{\lambda_i\}$ for each
set of $\{l(l+1), l_3, t_3\}$, $l=0,1,2,\cdots\cdots$, $-l\leq l_3\leq
l$, and they can be found by solving an ordinary differential
equation,
$$ 
\{-{\partial_x}^2 - {3\over x}{\partial_x} + {4l(l+1)\over
x^2}  -8 t_3l_3f(x)\}\Psi(x)=\lambda \Psi(x),
$$
where $\Psi$ is the ``radial" part of the eigenfunction. With $f(x)>0$, the
attractive case corresponds to
$l_3t_3=1$ and the
lowest energy level is $l=1$.  We find that all eigenvalues are
positive, except possibly one.  This zero eigenvalue solution can
be constructed by changing variable to $z=1/x^2$, and solving
\be
[-{d^2 \over d z^2}+V_{eff}(z)] \Psi = \lambda z^{-3} \Psi,
\ee
where $V_{eff}(z)=2/ z^2 - 2/z^3 f({1/\sqrt z}) = 2\rho^{-2}/(
z+\rho^{-2})$.  The zero eigenvalue solution is
\be
\Psi(x)=(1+\frac{1}{x^2})-(1+\frac{x^2}{2})\log(1+\frac{1}{x^2}).
\ee
The $x\rightarrow\infty$ limit is $\Psi(x)\rightarrow x^{-4}$ as it must.
But the $x\rightarrow 0$ limit is $\Psi(x) \rightarrow x^{-2}$, which
leads to logarithmic divergence in the norm at short distances. This
divergence seems to render this ``near'' zero mode questionable.
However, it should be recognized that our ``small loop'' solution
(\ref{eq:smallR}) is valid only for $R < x <\infty$. A proper treatment
of our small oscillations problem requires cutting out a small ball
around the origin, with a radius $r_0=O(R)$. Moreover, because of the
zero eigenvalue the divergence in the norm does not imply a divergence
in $G$.  We should in principle be able to solve this problem by find
the lowest eigenvalue and then letting $r_0$ go to zero. With respect
to this cutoff, our ``near'' zero mode is physically meaningful. An
alternative approach, which is under investigation, may be to use the
more general MA gauge with a Higgs potential. This has a natural
length scale $1/m_H$, which probably provides the necessary cutoff at
short distances.

As one extends beyond the quadratic approximation, one can see that
this zero mode leads to our ``exact solution'' in the special case of
$f=1/x^2$. That is, the formal $\rho\rightarrow \infty$
limit solution. In this sense, it is easy to see that the divergence
should be cut off by $\rho$. For the exact solution, the eigenvalue is
of the order $(R/\rho)^2 \log(\rho/R)$, which is consistent with our
numerical calculation as exhibited in  Fig. \ref{fig:R4beh}.

\section{Numerical Solutions to MA Projection}

For a general value of the monopole loop radius $R$, we were unable to
find an analytic solution. Consequently we have discretized our PDE's
and found numerical solutions. The need for a numerical integration
method is not surprising since even the 't Hooft-Polyakov monopole has
no known solution in closed form, except in the BPS limit.

With the simplifying ansatz (\ref{eq:ansatz}), we were able to reduce
the problem from a 4-d to a 2-d set of PDE's allowing us to construct
very accurate solutions on a 2-d grid. To check the validity of this
ansatz, we also consider the standard 4-d hypercubic grid
conventionally used in Monte Carlo studies of non-Abelian gauge field
theory.  An important advantage of the 4-d grid is the ability of
making a global search for stationary points. However this numerical
integration method should not be confused with Monte Carlo
simulations. Here the grid is used merely to solve numerically the
classical PDE's of Yang-Mills theory. As always for discretization
methods, it is important to consider carefully errors arising from the
grid spacing $a$ and the volume of the box. Our analysis of these
errors will shed some light on earlier investigations
\cite{Hart,Schierh,Markum} of the Abelian projection in ``cooled''
instanton configurations.

\subsection{Single Instanton Case}

Once we have made our ansatz (\ref{eq:ansatz}), $\beta(x,y,z,t)$ only
depends on two ``radial'' co-ordinates: $u = \sqrt{x^2 + y^2}$ and $v
= \sqrt{z^2 + t^2}$. This feature also generalizes to the case of the
MA projections~(\ref{eq:covglobal}) with a Higgs potential which
allows the magnitude $\phi(x,y,z,t) \equiv |\vec\phi|$ to
fluctuate. Here our ansatz (\ref{eq:ansatz}) implies that both
$\beta(x_\mu)$ and $\phi(x_\mu)$ depend only on $u$ and $v$. The
functional $G$ takes the simple from
\be
G = 4 \pi^2\int du dv \; u v \; [\half  M(\beta)\phi^2 + \half (\d_\mu
\phi)^2 + \quarter \lambda (\phi^2 - 1)^2 ] \; ,
\ee
where $\phi(u,v) = |\phi|$, $(\d_\mu \phi)^2 = (d\phi/du)^2 +
(d\phi/dv)^2$ and
$$
M(\beta) = 8 f(x^2) + (d\beta/du)^2 + (d\beta/dv)^2 + \sin^2\beta
(\frac{1}{u^2} + \frac{1}{v^2}) - 8 f(x^2) [ \sin^2 \beta -\half \sin
\beta \cos \beta (\frac{v}{u} - \frac{u}{v})].
$$
The conventional  MA gauge is given in  the limit $\lambda\rightarrow
\infty$. 

We introduce grids in u and v, mapping the infinite $u-v$ plane to a
finite region, for a range of values of $R/\rho$ and $\lambda$ (see
Fig.~\ref{fig:R4beh}). For all $\lambda$ (including the BPS limit
$\lambda =0$), we found that the monopole loop solution exists and
that the functional $G$ increases as $R^4$ within errors.  For all
values of $\lambda$ the global minima appeared to be at the point
where $R \rightarrow 0$ and the instanton returned to the singular
gauge. If we perturbed the background field slightly away from a pure
instanton by changing the functional form,
$f(x)=\rho^2/[x^2(x^2+\rho^2)]$, one can easily find field
configurations in the same topological sector which stabilized the
loop at a fix radius.  This strongly suggests that quantum
fluctuations or instanton interactions can cause monopole loop
formation.

To make a global search for minimal solutions for the MA projection
and to verify our theoretical ansatz, we have resorted to a 4-d grid.
A natural choice is the standard Wilson approach with link variables,
\be
U_\mu(x) = \E{i a A_\mu(x)} ,
\label{Ulink}
\ee
where $a$ is the lattice spacing. On this grid the MA functional
which leaves the Maximal Abelian subgroup $U(1)$ unconstrained is 
\be
G = \frac{1}{2} \; a^2 \; \sum_{\mu,x} \{ 1 - 
      \frac{1}{2}Tr( \hat \phi(x) U_\mu(x) \hat \phi(x+\mu)
                                  U_\mu(x)^{\dagger} )  \}.
\ee
In order to numerically approximate its value, we restrict the
summation over a finite volume V around the origin with open boundary
conditions on the surface. This is a good approximation for an
isolated instanton, since the contributions to the sum for large $x$
drops as $1/x^2$.  Furthermore to obtain a finite value for $G$ in the
infinite volume limit, we know that the gauge rotation must become a
constant at infinity. The minimization of the restricted functional
was done using the standard over-relaxation  algorithm \cite{Mandula}.

The instantons are placed on the grid in the non-singular gauge
and then rotated to singular form. Each link is approximated by
its integral using the trapezoid rule. The resulting configurations
gave topological charge and action deviating at worst by $10\%$ from
their continuum values, $1$ and $2\pi^2$ respectively.

We have investigated a range of different sizes of lattices and
instantons. As expected \cite{Hart,Schierh}, we have found monopole loop
formation in the Maximal Abelian gauge. But our monopole radii do not scale
with the instanton size. This is a clear signal that the
stabilization of the monopole loop is a lattice artifact.  Despite
that, we find that once a loop is formed the gauge rotation accurately
satisfies our theoretical ansatz (\ref{eq:ansatz}) that $\alpha
=\varphi-\psi$ and $\beta$ depends only on $u$ and $v$.

\begin{figure}
$$
\epsfxsize=8.6cm
\epsfbox{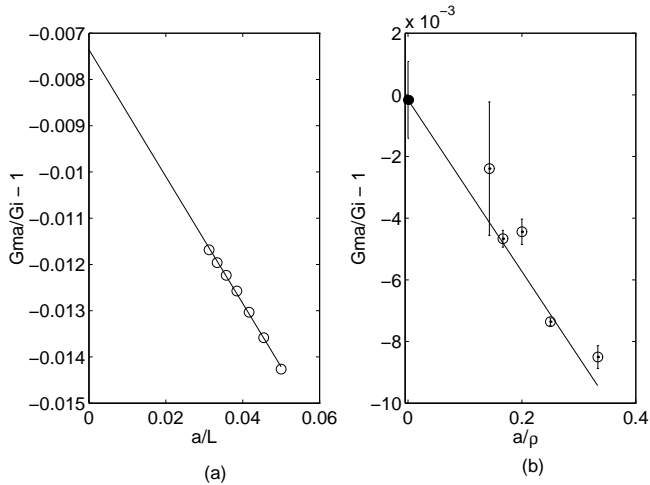} 
$$
\caption{ The finite lattice spacing and finite volume dependence for the
monopole loop solution on the lattice.(a) Typical Infinite volume 
extrapolation.(b) $a \rightarrow 0 $ extrapolation.
\label{fig:Scaling} }
\end{figure}

A finite volume and finite spacing analysis has been  done in order to
further establish the fact that the monopole loops is a lattice artifact.
We measure the change, 
\be
\Delta(\rho/a,L) = \frac{ G_{MA} - G_{instanton}}{G_{instanton} } \; ,
\ee
of the MA functional after gauge fixing and we demonstrate that it
vanishes in the combined limit of infinite volume and zero lattice
spacing.  In order to show this, we have found a series of solutions for
lattice sizes $ 20^4 $, $ 22^4 $, $ 24^4 $, $ 26^4 $, $ 28^4 $, $ 30^4
$ and $32^4$ with a fixed instanton size. The infinite volume limit is
obtained by doing a linear extrapolation in $1/L$ , where $L=V^{1/4}$
is the linear dimension of our lattice. Fig. \ref{fig:Scaling}(a) shows
this extrapolation for $\rho/a = 4 $.  The second extrapolation 
to zero lattice spacing is done from these extrapolated values of
$\Delta$ for various values of the instanton radius $\rho$. 
At infinite volume, the only scale in the problem is $\rho$, and it is 
therefore defines the lattice spacing, $a\sim 1/\rho$. Thus one takes 
the $a \rightarrow 0 $ limit by taking the $\rho/a \rightarrow \infty $ 
limit. This extrapolation is done using instanton sizes 
$\rho/a = 3, 4, 5, 6, 7$. 

The results are shown in Fig. \ref{fig:Scaling}(b).  The extrapolated
value of $\Delta(\rho/a,L)$ in the limit of $a \rightarrow 0$ and $L
\rightarrow \infty $ is clearly zero within the numerical error. 
Another indication that the global minimum of the MA functional is the
instanton in the singular gauge is the fact that $\Delta$ is so small
$O(10^{-3})$. This is also supported by our experience with the
minimizing routine. Starting with the gauge background for an
instanton in the singular gauge, it took very few iterations to
converge which shows that the displacement of the minimum due to
lattice artifacts is actually very small.
 
\subsection{Solutions for Instanton Pairs.}

We have also used our 4-d grid to begin to investigate solutions to the MA
projection in the background of interacting instanton pairs.  For
example, we have begun to study instantons (I-I) pairs and instanton
anti-instantons pairs (I-A) with various sizes, relative separations
and relative isospin orientations.  Already several general conclusions
can be drawn.  The I-I pairs appear to be very similar to the
case of single instantons. Indeed for the 't Hooft ansatz, one can
again show that the two instanton solutions (or indeed the
multi-instanton solution),
\be
\vec A_\mu\cdot\vec\tau =\tau^\alpha \bar\eta^\alpha_{\mu\nu} \d_\nu ln( 1 +
\rho_1^2/x_1^2 + \rho_2^2/x_2^2), 
\ee
already satisfies the MA gauge condition.  On the 4-d grid our
preliminary study indicates that the MA projection of the two
instantons have small loops centered at each instanton which again
appear to be due entirely to lattice artifacts.  It is natural to ask
whether there is a general property of all exact classical solutions
that the global minimum of the MA projection has no monopole
trajectories.

\begin{figure}
$$
\epsfxsize=8.6cm
\epsfbox{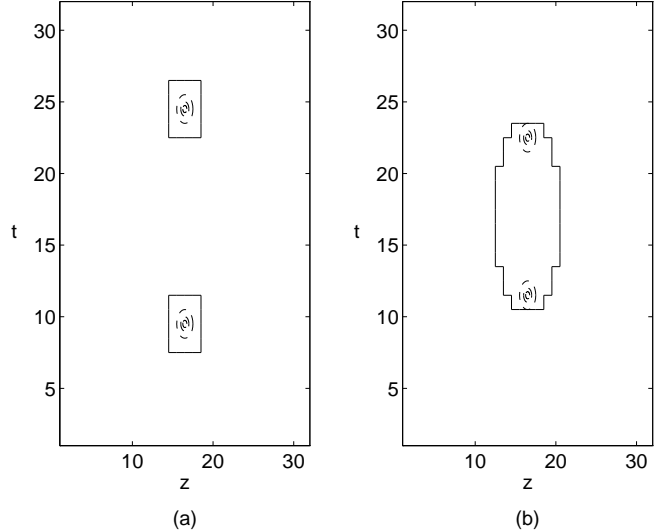} 
$$
\caption{ Monopole loops for I-A molecule with the most
attractive orientation: (a) Loops stabilized at separation $d/\rho = 1.88$.
(b) Loops fusion at separation $d/\rho = 1.38$.
\label{fig:Interaction} }
\end{figure}

On the other hand, we have computed the MA projection for I-A pair
with isospins oriented in the most attractive configuration.  In this
case, contrary to the single instanton and I-I pair large loops are
formed in the MA projection. At large separation, each instanton in
the I-A pair has its own monopole loop as illustrated in
Fig~\ref{fig:Interaction}, but as the separation $d$ is decreased the
individual loops fuse at $d/\rho \simeq 1.88$ to form a large loop
which surrounds the I-A pair. These loops clearly scale with the size
of the system and the reduction of the MA functional is clearly
larger. Further more this reduction in $G$ becomes larger as you
approach the continuum limit. Although we have not yet finished the
same detailed finite size and finite lattice spacing analysis as in
the single instanton case, we have considerable evidence that this
effect will survive in the continuum.

In the dipole interactions between I-A pairs, we believe we are seeing
the first semi-classical mechanism for nucleating monopole loops.  As
the instantons become denser, the monopole loops begin to percolate
between the individual instantons.  It is easy to imagine that in an
instanton liquid there is a critical density at which the percolation
clusters are infinite and the monopoles can be said to condense. This
effect is an intriguing possibility for a semi-classical mechanism for
confinement. Additional studies of monopole trajectories for
interacting instantons and further analysis to support (or refute)
this scenario are postponed to a future publication~\cite{BOT}.

\section{Conclusions}

The main goal of this paper was to find the earliest point in the
semi-classical instanton vacuum in which the 't Hooft Mandelstam monopole
appears. We believe that this is the formation of small current loops
centered at each instanton and anti-instanton.  In the extreme limit
of infinite separation, i.e.,  an isolated instanton, the loops shrink
to zero. However to $O(R^4\log R)$ in the radius there is a flat direction
for loop formation in the gauge fixing functional. We have begun to
investigate the effect of  instanton anti-instanton interactions. Here
it appears that when the relative isospin orientation of an instanton
anti-instanton (I-A) pair is most attractive, the monopole loops are
stabilized at a finite radius and as the pair moves closer together
the individual loops fuse. 

There is much more to learn about the mechanism for loop
stabilization and the effects of instanton interactions. For the most
part we postpone this to a future publication~\cite{BOT}. However it
is worth posing some of the questions and extending several arguments
touched on early.

First we have noticed that the single instanton in the singular gauge
already satisfies the differential form of the MA gauge and
gives rise to a finite contribution to the gauge fixing functional $G$.
In fact it is trivial to see that all multi-instanton configurations
that satisfy the 't Hooft ansatz $A^a_\mu =  \bar\eta^a_{\mu\nu} \d_\nu 
ln( 1 + F(x^2))$ likewise satisfy the MA gauge with a finite
contribution for $G$. Thus it is tempting to conjecture that
any exact self-dual classical solution minimizes this functional
without explicit monopole currents. We are investigating this
conjecture further. In this scenario the essential mechanism for
monopole loop formation would be the interaction terms
between self-dual and anti-self-dual regions that act as the ``domain walls''
between I-A pairs.

Next it is worth expanding a little on the relationship between the
topological charge and the monopole charge for an
isolated instanton. In the singular gauge, when we write the
topological charge, $Q = (1/16 \pi^2) \int d^4x Tr[ \tilde F_{\mu
\nu} F_{\mu \nu}]$, in terms of the gauge variant current, $Q =
\int d^4x \d_\mu K_\mu$, one must exclude singular regions. Indeed
via Gauss' theorem, the flux into these excluded singularities gives
the net topological charge. We have studied carefully how this
theorem is satisfied in the presence of our monopole loop.
In the singular gauge, we may write the
instanton as
\be
A_\mu = \bar A_\mu + M_\mu = - \frac{ g^\dagger \d_\mu g}{i} \frac{x^2}{ x^2 +
\rho^2} + 
\frac{g^\dagger \d_\mu g}{i}.
\ee
As we shrink a small sphere toward the origin $x^2 = \delta^2
\rightarrow 0$ the surface area is $O(\delta^3)$. However one sees that the
singular part of the current,
\be
K_\mu =  \frac{1}{8 \pi^2} \epsilon_{\mu \nu \rho \lambda} Tr[ A_\nu (\d_\rho
A_\lambda -  i
\smfrac{2}{3} A_\rho A_\lambda)],
\ee
diverges as $O(1/\delta^3)$ and that the contribution to $Q$ comes
entirely from the pure gauge piece, $M_\mu$. On the other hand as we
rotate the field configuration into a monopole loop,
\be
M_\mu \rightarrow M_\mu =  \frac{\Omega g^\dagger \d_\mu (\Omega
g^\dagger)^\dagger}{i} \; ,
\ee
the singularity is spread out to the loop at radius $x^2 =
R^2$. Now the toroidal surface area around the loop is $O(R
\delta^2)$, but again a careful analysis shows that the only divergent
piece of $O(1/\delta^2)$ comes from $M_\mu$.  The argument is
completely general for a monopole loop gauge and it requires the
magnetic charge to be quantized to match the topological charge.  This
shows that the monopole loop has a very natural fit to the instanton.

Finally we have found that when an isospin orientation of an isolated
instanton is fixed, the monopole loops are constrained to a portion of
the full $O(4)$ group.  Again there is a rather remarkable
``coincidence'' this time between the symmetries of the monopole loop
and the instanton. A single loop ``kinematically'' breaks the Lorentz
group $SU_L(2) \times SU_R(2)$ into a factor of coset spaces
$SU_L(2)/U_L(1) \times SU_R(2)/U_R(1)$ in the same sense that the
't~Hooft-Polyakov monopole brakes the $SU(2)$ isospin symmetry down to
$SU(2)/U(1)$. Then if we fix the isospin axis of the instanton
(anti-instanton) and choose the MA gauge relative to $\tau_3$, the
loop is only re-oriented by the right (or left) coset. This implies a
correlation between the plane of the loop and the isospin orientation,
which can be tested in typical background configurations of instantons
generated in a Monte Carlo simulation. In this way we can determine
whether or not our conjecture that these monopole loops are important
configurations for the full quantum theory is correct.

In conclusion, although there are many more details worth considering
there is a remarkable coincidence between the form of an instanton
and its monopole loop in the MA projection. This is reflected in
topological, symmetry and stability terms. This leads us to see the
instanton in a new light as the ``seed'' for the formation of monopole
loops. The dynamical implications are much more difficult, but it
appears that I-A interactions may play a crucial role and the large
``entropy'' of monopole loops percolating between near by instantons
suggest a promising direction for future research on electric
confinement.

\section*{Acknowledgments}
 We gratefully acknowledge important help and
encouragement by U-J Wiese and very helpful conversations with M.
Chernodub, F. Gubarev and M. Polikarpov. One of us (RCB)
gratefully acknowledges the warm hospitality of the Center for
Theoretical Physics at MIT. Two of us (RCB and CIT) would also like to thank
the Aspen Center for Physics for their participation in the 1996 summer
program. The computational work in support of  this research was performed at
the Theoretical Physics Computing  Facility at Brown University. This work is
supported in part  by the D.O.E.  Grant
\#DE-FG02-91ER400688, Task A.

\appendix
\section{Gauge Covariant Formulation of the Abelian projection}

The usual way to study the MA gauge is to applying the
gauge transformation to the field,
\be
A_\mu(x) \rightarrow A^\Omega_\mu(x) 
= \Omega(x) A_\mu(x)\Omega^\dagger(x) \; + \;  \frac{1}{ i e}
\Omega(x) \d_\mu \Omega^\dagger(x) \; .
\label{eq:Decomp}
\ee
and find that rotation $\Omega(x)$ which satisfies a gauge condition.
Alternatively, one may introduce an auxiliary adjoint Higgs-like field,
$\Phi = \vec \phi \cdot \vec \tau$, expressing the entire problem in a
gauge invariant form. We refer to this two methods as ``active'' and
``passive'' respectively. In this appendix, we collect together basic
formalism for each form. 

In the passive description, the MA gauge is cast in a gauge covariant
form familiar to the Higgs~\cite{KSLW} model, with the entire
formalism re-expressed {\bf exactly} in the form used to construct the
't Hooft-Polyakov monopole.  This covariant formulation both
simplifies our analysis and emphasizes the important physical point
that the Maximal Abelian projection needs not be viewed as a gauge
fixing prescription but rather as a way to define magnetic degrees of
freedom. We prefer the latter interpretation.\footnote{There is also
a considerable literature \cite{Polikarpov,Smit}, which goes on to try
to construct an appropriate monopole condensate order parameter in the
$U(1)$ sector. Our field $\Phi$ is not this object, since it, like the
Higgs field, lives in the coset space.}

\subsection{Active View of MA Gauge and Magnetic Current} 
 
In the active form, the resulting Abelian field $a_\mu(x)= Tr[\tau_3
A^\Omega_\mu(x)]$, (or more properly its derivatives), which depends
on the choice on the non-Abelian background field, can have
singularities leading to a non-zero magnetic current.  Introducing the
notation, $\bar A_\mu(x) \equiv \Omega(x) A_\mu(x)\Omega^\dagger(x)$
and $M_\mu(x) \equiv (1/ i e) \Omega(x) \d_\mu \Omega^\dagger(x)$, the
transformed field becomes
\be
A^\Omega_\mu(x) = \bar A_\mu(x) \; + \;M_\mu(x).
\label{eq:DecompTwo}
\ee 
This
splits the Abelian field into two components, $a_\mu(x) = \bar
A^3_\mu(x)+M^3_\mu(x)$. The first term, $\bar A_\mu^3$, for our
problem, comes from a direct rotation of the instanton field. The
second ``induced'' term, $M^3_\mu$, can contain monopoles as its
source when appropriate conditions are met as we demonstrate next.

The Abelian field strength, $f_{\mu\nu} \equiv \d_\mu a_\nu -
\d_\nu a_\mu$, is given by 
\be
f_{\mu\nu} = (\Omega F_{\mu\nu} \Omega^\dagger)_3
- i e [\Omega (A_\mu + \frac{1}{ i e} \d_\mu) \Omega^\dagger , 
\Omega (A_\nu + \frac{1}{ i e} \d_\nu )\Omega^\dagger]_3
\label{eq:fmunu1}
\ee
where $(\cdots)_3 =  Tr[\tau_3 \cdots]$. It again is  split into two pieces,
\be
f_{\mu\nu} = (\d_\mu \bar A^3_\nu - \d_\nu \bar A^3_\mu)
+ (\d_\mu M^3_\nu - \d_\nu M^3_\mu) .
\label{eq:fmunu2}
\ee
Since the dual of the first combination is obviously divergentless, only
the second term contributes to a non-vanishing  magnetic current,
\be
k_\mu(x) = \frac{e}{8 \pi }\epsilon_{\mu\nu \rho \sigma} Tr(\tau_3  \d_\nu  
[M_\rho(x), M_\sigma(x)]),
\label{eq:MAcurrent}
\ee
where $k_\mu = \smfrac{1}{4\pi} \d_\nu \tilde f_{\nu\mu}$, and $ \tilde
f_{\nu\mu}= \half \epsilon_{\mu\nu\rho\sigma}f_{\rho\sigma}$.

\subsection{Gauge Invariant View of MA Projection}

We now reformulate the MA projection in passive form.  Let us begin by
noting that the functional,
\be
G = {\quarter} \int (A^1_\mu + i A^2_\mu)(A^1_\mu - i A^2_\mu) d^4x 
\ee
is just the mass term in the broken phase of an $SU(2)$ Georgi-Glashow
model.  This suggests a change of variables from $\Omega(x)$ to
\be
\Phi(x) = \Omega^\dagger(x) \tau_3 \Omega(x).
\ee

The functional $G$ now takes form of
action for a Higgs field,
\be
G =  \half \int d^4x\{ \half [D_\mu(A) \vec\phi]^2 + V(\vec\phi^2)\},
\ee
where $\Phi=\vec\phi\cdot\vec\tau$, $D_\mu\phi^a\equiv
\d_\mu\phi^a+e\epsilon^{abc}A_\mu^b\phi^c$, and  the potential,
\be
V(\vec\phi^2)=\sigma (\vec\phi^2-1),
\ee
involves a Lagrange multiplier $\sigma $ in order to maintain the constraint
$\vec\phi\cdot\vec\phi =1$. We also suggest a generalization of the MA
projection to include the standard quartic Higgs potential,
\be
V(\vec\phi^2)= \quarter \lambda (\vec \phi^2 - v^2)^2, 
\ee
with $\Phi(x)$ now given by $\vec \phi(x)\cdot
\vec\tau/|\vec\phi(x)|$. This more general from we will refer to as
the ``Higgs MA projection''. Thus the MA gauge is precisely the same
as minimizing the action for a non-dynamical (or auxiliary) Higgs-like
field in a background Yang-Mills theory. The limit $\lambda
\rightarrow \infty$ at fixed VEV $v$ results in the usual MA
projection. This is the limit which takes the Higgs mass $m_H$ to
infinity. This formulation corresponds to a passive description of a
gauge transformation. In the active description, rotation $\Omega$ is
applied directly to the gauge field in Eq. (\ref{eq:global}) so that
$\hat \phi \rightarrow (0,0,1)$.  Clearly, the passive and the active
descriptions are equivalent: each specifies a gauge transformation
$\Omega$ up to the $U(1)$ subgroup. However the passive description is
more ``natural'' since $\hat \phi$ lives in the coset space.

The Abelian field strength (\ref{eq:fmunu1}) now takes a manifestly
gauge invariant form,
\be
f_{\mu \nu} =  \hat\phi^a F^a_{\mu\nu} - i e \epsilon^{abc} \hat \phi^a
D_\mu \hat \phi^b D_\nu \hat \phi^c ,
\label{eq:gauinvF1}
\ee
and the magnetic current is the well known
conserved  topological current,
\be
k_\mu(x) = \frac{1}{8  \pi e}\epsilon_{\mu\nu \rho \sigma} \epsilon^{abc}
\d_\nu \hat \phi^a \d_\rho \hat \phi^b \d_\sigma \hat \phi^c.
\label{eq:HiggsCurrent}
\ee
The essential identity in deriving these expressions from our earlier equations
is
$\d_\mu \Phi = i [ \Omega^\dagger M_\mu \Omega, \Phi]$. 
An alternative form for $f_{\mu\nu}$ follows directly from Eq.
(\ref{eq:fmunu2}),
\be
f_{\mu \nu} = \d_\mu \bar A^3_\nu -  \d_\mu \bar A^3_\nu  - i e \epsilon^{abc}
\hat\phi^a
\d_\mu \hat\phi^b \d_\nu \hat\phi^c .
\label{eq:gauinvF2}
\ee
This approach to MA projection not only provides a gauge invariant
treatment, but also allows a topological interpretation for the
possible presence of magnetic sources. For example, as in the case of
the 't Hooft-Polyakov monopole, the presence of a monopole charge can
be understood to be due to a non-trivial homotopy $\Pi_2(SU(2)/U(1)) =
Z$.
 
\subsection{ Higgs MA Projection for Instanton}

For general reference to our analysis, we final give the full set of 
differential equation for the Higgs MA projection,
\be
D_\mu(A)^2\vec \phi - \vec \nabla_{\phi} V(\phi)  = 0 \; .
\label{eq:HiggsMA}
\ee
The two independent PDE's for the tangential components,
now take the form
\bea
 \phi \; \d^2_\mu \alpha +2\d_\mu\phi\d_\mu\alpha +2\phi\; \cot \beta \;
(\d_\mu\alpha)(\d_\mu\beta) = -2 [\phi (
\hat\phi \cdot
\vec A_{\mu})\; ( \d_\mu\beta)-(\hat\beta\cdot\vec A_\mu)\;(\d_\mu\phi)],
\label{eq:MAGenerala}\\
\phi\d^2_\mu \beta+2\d_\mu\phi\d_\mu\beta -\half \phi\;\sin 2\beta\;
(\d_\mu\alpha)^2 = 2[\phi\sin\beta (\hat \phi
\cdot
\vec A_{\mu}) \;  (\d_\mu\alpha )-(\hat\alpha\cdot \vec A_\mu)(\d_\mu\phi)],
\label{eq:MAGeneralb}
\eea
by projecting Eq.~(\ref {eq:MAgauge}) onto the $\hat\alpha(x)$ and $\hat
\beta(x)$ axes respectively.  In addition there is a new
equation for the radial mode,
\be
\d_\mu^2\phi=-\{2[\sin\beta(\hat\beta\cdot\vec
A_\mu)(\d_\mu\alpha)+(\hat\alpha\cdot\vec A_\mu)\d_\mu\beta]+[V'(\phi^2) -
\smfrac{2}{3} \vec A^2_\mu]\}\phi.
\label{eq:MAGeneralc}
\ee

Under our ansatz, $\alpha=\varphi-\psi$, with $\beta(u,v)$ and
$\phi(u,v)$ only functions of $u$ and $v$, Eq.~(\ref{eq:MAGenerala})
is still satisfied automatically.  Now Eq.~(\ref{eq:MAGeneralb}) 
becomes
$$
\phi\d^2_\mu \beta+2(\d_u\phi\d_u\beta+\d_v\phi\d_v\beta) -
(\frac{u^2+v^2}{2u^2v^2})
\phi\sin(2\beta)
=
$$
\be
 2 f(x)\{\phi [(\frac{v^2-u^2}{uv})\sin^2\beta-\sin
2\beta]+[v\d_u\phi-u\d_v\phi]\},
\ee
 and the the radial equation
(\ref{eq:MAGeneralc}) becomes
\be
\d_\mu^2\phi=-\{ f(x)[\sin
2\beta(\frac{v^2-u^2}{uv})+4\sin^2\beta+2(u\d_v\beta-v\d_u\beta)] +[V'(\phi^2)
-\smfrac{2}{3}\vec A^2_\mu]\}\phi.
\ee
Both analytical and numerical properties of these equations are
discussed in the text. There appears to be remarkably little
dependence of our monopole loop solution on $\lambda$ from $\lambda =
0$ (the BPS limit) to $\lambda = \infty$ (the standard MA projection).

\subsection{Abelian Projected Theory}

In the MA gauge, one is left with an intermediate description of a
$U(1)$ gauge theory, 
\be
{\cal L}(a_\mu, A^{\pm}_\mu) = \quarter f_{\mu\nu}^2 + (d_\mu A^+_\nu)( d_\mu
A^-_\nu) +  e^2 (A^+_\mu  A^-_\nu - A^+_\nu  A^-_\mu)  A^+_\mu  A^-_\nu. 
\ee
interacting exclusively through charged vectors $A^{\pm} = A^1_\mu \pm
i A^2_\mu$.  The $U(1)$ covariant derivative is $d_\mu = \d_\mu + i e
a_\mu$.  

In addition to the standard Wilson loop $W({\cal C}) = Tr{\cal P}_{\cal
C} \E{ i \int dx_\mu A_\mu}$, one may introduce the Abelian Wilson
loop,
\be
W^{ab}({\cal C}) = \E{ i \int dx_\mu a_\mu}
\ee
which in turn can be split up into ``photon'' and ``monopole'' factors
\be
W^{ab}({\cal C}) = \exp[\;i \int dx_\mu \bar A^3_\mu \;]  \;\;\; \exp [ i  e
\int\int d\sigma_{\mu\nu}
\epsilon^{abc} \hat \phi^a \d_\mu  \hat \phi^b \d_\nu  \hat \phi^c].
\ee
The second integral is taken over any surface whose boundary is given
by the loop ${\cal C}$. The near ``saturation'' of the non-Abelian
string tension by the Abelian part in the MA gauge, $\sigma_{ab} \simeq .92
\sigma$, and the Abelian part by the monopole contribution, $\sigma_{monopole}
\simeq .95 \sigma_{ab}$, observed in lattice simulations \cite{Schilling} is
referred to as Abelian Dominance.  If this survives the continuum
limit this may provide the dynamical link between monopole
configuration and confinement.

To understand the role that Abelian dominance might play in the
continuum theory, it is useful to look at the very interesting form of
the Wilson loop suggested by Diakonov and Petrov \cite{DP},
\be
W({\cal C}) = 
\int {\cal D}\hat \phi(x) \exp[ i\half \int dx_\mu \hat \phi \cdot \vec A_\mu]
\;\;\;  \exp [ \int d\sigma_{\mu\nu} \epsilon^{abc} \hat \phi^a
\d_\mu \hat \phi^b \d_\nu  \hat \phi^c].
\ee
A comparison with the Abelian Wilson loop in the MA gauge shows that
this has {\bf exactly} the same form with the additional step that one
must average over all gauge transformation in the $SU(2)/U(1)$
coset. Consequently Abelian dominance is the statement that in the
true quantum vacuum, the contributions to the $\hat \phi$ average is
approximated by the MA projection. Thus one strategy to proving
confinement is to first establish Abelian dominance (or more precisely
the inequality $W({\cal C}) \le W^{ab}({\cal C})$ for large loops) and
then to demonstrate that monopole condensation occurs forcing an area
law for the Abelian loop.


\begin{references}

\bibitem{tHooft2}  G. 't Hooft, Phys. Rev. Lett. 37, 8 (1976); 
                   Phys. Rev D14, 3432 (1976).

\bibitem{Shuryak} E.V. Shuryak, Nucl. Phys. B302, 559-644 (1988).

\bibitem{Witten} E. Witten, Nucl. Phys. B156, 269 (1979); 
                 G. Veneziano, Nucl. Phys. B159, 213 (1979).

\bibitem{Grandy} M.-C. Chu, J.M. Grandy, S. Huang, J.W. Negele, 
                Phys. Rev. D49, 6093 (1994); Phys. Rev. D48, 3340 (1993).

\bibitem{Mand} S. Mandelstam, Phys. Rep. C23, 245 (1976).

\bibitem{tHooft1} G. 't Hooft, Nucl. Phys. B190 (1981) 455.

\bibitem{Peskin} M. E. Peskin , Ann. of Phys. 113, 122 (1978). 

\bibitem{Poly} A. Polyakov Phys. Lett. B59, 82 (1975); 
               Nucl. Phys. B120, 429 (1977).

\bibitem{Seiberg-Witten} N. Seiberg and E. Witten, Nucl. Phys. B426, 19 (1994).

\bibitem{PolikarpovRev} M. I. Polikarpov, Lattice 96 Review talk
hep-lat/9609020 (1996).

\bibitem{BPST} A. A. Belavin, A. M. Polyakov, A. S. Schwartz and 
Tyupkin, Phys. Lett. B59, 85 (1975); M. F.  Attyah, N. J. Hitchin,
V. G. Drinfeld and Yu. I. Manin,  Phys. Lett. A65, 185 (1978).
                 
\bibitem{Tanaka} H. Suganuma, A. Tanaka, S. Sasaki, O. Miyamura,
                 Nucl. Phys. B(Proc. Supp.), Proc. of Lattice 95;
                 H. Suganuma, K. Itakura, H. Toki, HEPTH-9512141,
                  (hep-th/9512141).

\bibitem{Rossi} P. Rossi, Nucl. Phys. B149, 170 (1979).

\bibitem{CG} M.N. Chernodub and F.V. Gubarev ITEP-95-34, hep-th/9506026.

\bibitem{Hart} A. Hart and M. Teper Oxford preprint No: OUTP-95-44-P,
              hep-lat/9511016.

\bibitem{Schierh} V. Bornyakov, G. Schierholz DESY 96-069,
                  HLRZ 96-22, hep-lat/9605019 ;
                  G. Schierholz DESY 95-127 HLRZ 95-35.

\bibitem{Markum} H. Markum, W.Sakuler, S. Thurner, Nucl. Phys B
               (Proc. Supl. LATTICE95 ), 254 (1995) (hep-lat/9510024);
               S. Thurner, M. Feurstein, H. Markum, W. Sakuler,
               Phys. Rev D54 3457(1996). 

\bibitem{BOT} R. C. Brower, K. Orginos and C-I Tan,
              ``Monopole loops for interacting instantons", (in preparation).

\bibitem{KSW} A. S. Kronfeld, G. Schierholz and Wiese, 
             Nucl. Phys. B293, 461 (1987).

\bibitem{Rajaraman} R. Rajaraman, {\it Solitons and Instantons},  
                   (North-Holland Publishing Company, 1982).

\bibitem{Belavin} A. Belavin, A. Polyakov, A. Schwartz and Y. Tyupkin, 
                 Phys. Lett. B59, 85 (1975).

\bibitem{Jakiw} R. Jackiw, C. Nohl, C. Rebbi, Phys. Rev. D15 No 6, 1642 (1977).

\bibitem{Dirac-Sheet} M. Blagojevi'c and P. Senjanovi'c,
                     Phys. Rep. 157, Nos 4 \& 5, 233 (1988).

\bibitem{Mandula} J. E. Mandula, M. Oglivie, Phys. Lett. B248, 156 (1990).

\bibitem{KSLW} A. S. Kronfeld, M. L. Laursen, G. Schierholz and Wiese,
              Phys. Lett.  B198 No 4, 516 (1987).

\bibitem{Polikarpov} M.N. Chernodub, M.I. Polikarpov, A.I. Veselov,
                    Ahrenshoop Symp. 307 (1995) (hep-lat/9512030);
                    T.L. Ivanenko, A.V. Pochinsky, M.I. Polikarpov,
                    Nucl. Phys. B, (Proc. Suppl. 30 ), 565 (1993);
                    M.N. Chernodub, M.I. Polikarpov, A.I. Veselov,
                    Trento QCD Workshop, 81 (1995) (hep-lat/9512008).

\bibitem{Smit} J. Smit and A. J. van der Sijsm Nucl. Phys. B355, 603 (1991).

\bibitem{Schilling} G. S. Bali, V. Bornyakov, M. Mueller-Preussker, and K.
Schilling, hep-lat/960312, to be published in Phys. Rev. D.
 
\bibitem{DP} D.I.Diakonov and V.Yu. Petrov, Phys. Lett. B 224 (1989) 131.

\end{references}
\end{document}